\documentclass[preprint,showkeys,secnumarabic,amsfonts,showpacs,amsmath,amssymb]{revtex4}
\usepackage[dvips]{color}
\usepackage{array}
\usepackage{rotating}
\usepackage{epsfig}
\usepackage{amsmath}
\usepackage{graphicx}
\begin{document}
\title{Searching for  Cosmological Preferred Axis using  cosmographic approach}

\author{Amin.Salehi}\author{Mohammad Reza Setare}
\email{salehi.a@lu.ac.ir} \affiliation{Department of Physics,   Lorestan University , Lorestan, Iran}
\email{rezakord@ipm.ir} \affiliation{Department of Science, Campus of Bijar, University of Kurdistan, Bijar, Iran.}

\begin{abstract}
 \noindent \hspace{0.35cm}
 Recent released Planck data and other astronomical observations show
that the universe may be anisotropic on large scales. This hints a cosmological privileged axis in our anisotropic expanding universe. This paper proceeds a modified redshift in anisotropic cosmological model as $ 1+\tilde{z}(t,\hat{\textbf{p}})=\frac{a(t_{0)}}{a(t)}(1-A(\hat{\textbf{n}}.\hat{\textbf{p}}))$ (where $A$ is the magnitude of anisotropy ,$\hat{\textbf{n}}$ is the direction of privileged axis, and $\hat{\textbf{p}}$ is the direction of each SNe Ia sample to galactic coordinates) along with anisotropic parameter $\delta=\frac{A(\hat{\textbf{n}}.\hat{\textbf{p}})}{1+A(\hat{\textbf{n}}.\hat{\textbf{p}})}$. The luminosity distance is expanded with model-independent cosmographic parameters as a function of modified redshift $\tilde{z}$. As the  transformation matrix $M(n\times n)$ is obtained to convert the Taylor series coefficients of isotropic  luminosity distance  to   corresponding anisotropic parameters. These results culminate the magnitude of anisotropy  about $\mid A\mid \simeq 10^{-3}$ and the direction of preferred axis as $(l,b)=(297^{-34}_{+34},3^{-28}_{+28})$, which are  consistent with other studies in $1-\sigma$ confidence level.

\end{abstract}


\maketitle
\newpage

\section{Introduction\label{Int}}
According to cosmological principle, as one of the basic assumptions of modern cosmology, universe is homogeneous and
isotropic on scales larger than a few hundred Mpc. Indeed, this assumption
 is consistent with former astronomical observations such as the Cosmic Microwave Background (CMB) radiation data from the Wilkinson Microwave Anisotropy Probe (WMAP) \cite{Komats}-\cite{Hinshaw}, and data from type Ia supernovae, such as
those collected in the so-called Union \cite{Kowalski} , Union2 \cite{Amanullah}
compilations . However,  some challenges are reported to the cosmological principle  in recent years. For example, the large-scale alignments of the
quasar polarization vectors \cite{Hutsemekers}-\cite{Hutsemekers1},the large-scale bulk flow \cite{Kashlinsk}-\cite{Lavaux}, the alignments of low multipoles in CMB angular power spectrum\cite{Lineweaver}-\cite{Frommert},  the spatial variation of fine-structure constant
\cite{Dzuba}-\cite{King}, the CMB hemispherical asymmetry observed by WMAP \cite{Bennett1}-\cite{Bennett2} and
Planck satellite \cite{Ade}, Dark Energy Dipole\cite{a35}-\cite{a128},
 and other effects \cite{s18}-\cite{Perivolaropoulos}. \\

  There are many models, in the theoretical aspect, which have been proposed to resolve the asymmetric anomaly of these data. Several groups  have applied
the hemisphere comparison method to study the
anisotropy of $\Lambda CDM$, $wCDM$ and the dark energy model
with $CPL$ parametrization. So that, the supernova data
and the corresponding cosmic accelerations on several
pairs of opposite hemispheres have been used to search
for maximally asymmetric pair and a statistically significant
preferred axis \cite{Schwarz}.
Ref \cite{Antoniou} have applied the hemisphere comparison method
to the standard $ \Lambda CDM$ model and found that the hemisphere of maximum accelerating expansion is in the direction $(l; b) =309^{-3}_{+23},18^{-10}_{+11} ) $ with Union2 data. Also the Ref \cite{Kalus} tests the isotropy of the expansion of the Universe by estimating the hemispherical anisotropy of supernova type Ia (SN Ia)
Hubble diagrams at low redshifts $(z < 0.2)$ and found
a maximal hemispheric asymmetry towards a direction close to the equatorial
poles with Hubble anisotropy of $ \frac{\Delta H}{H}= 0.026$. The anisotropic cosmological model is investigated by Ref.\cite{chang1} in the Randers spacetime,  to determine the privilege direction of  $(l, b) =306 ,-18 ) $.
\\
\begin{table}[ht]
\caption{Incomplete list of pervious studies} % % title of Table
\centering % used for centering table
\begin{tabular}{|c |c | c| c|} % centered columns (8 columns)
\hline %inserts double horizontal lines
1  &$models$  &  $Equation$ &  $Ref$\ \\ [2ex] % inserts table
%heading
\hline %inserts single line
1  &scalar perturbation &($d_{L}=(1+z)\frac{c}{H_{0}}\int_{0}^{z}\frac{(1-d\cos\theta)dz}{\sqrt{\Omega_{m0}(1+z)^{3}+1-\Omega_{m0}
-\frac{4d\cos\theta(1+x)^{5}}{3H_{0}^{2}d^{2}_{L0}}}}$) &  \cite{Li},\cite{Wang} \ \\
\hline % inserts single horizontal line
2  &Anisotropic $d_{L}$ in the Finslerian space-time &($d_{L}=(1+z)\frac{c}{H_{0}}\int_{0}^{z}\frac{(1-d\cos\theta)^{-1}dz}
{\sqrt{\Omega_{m}(\frac{1-d\cos\theta}{1+z})^{-3}+\Omega_{\Lambda}}}$) & \cite {chang1} \\
\hline %inserts single line
3  & effect of peculiar velocities on$d_{L}$  \ &$\frac{\Delta d_{L}}{d_{L}}=\hat{n}.[\vec{v}_{pec}-(\vec{v}_{pec}-\vec{v}_{obs}).\frac{(1+z)^{2}}{H(z)d_{L}}]$&  \cite{Hui1}- \cite{Bonvin}\\
\hline %inserts single line
4  & "wind" scenario to the bulk flow&($d_{L}=(1+z)\int_{0}^{t}\frac{dt'}{a(t')}(1+d \cos\theta)=\bar{d_{L}}(1+d \cos\theta)$)\ &  \cite{Li3}\\
\hline %inserts single line
5  & dipole+monopole fit approach &($\frac{\Delta\mu}{\mu}=d cos\theta +m$) &   \cite{Xiaofeng},\cite{Antoniou},\cite{Wang}\ \\
\hline %inserts single line
6  &hemisphere comparison method&$\frac{\Delta\Omega_{0m}}{\bar{\Omega_{0m}}}=
2(\frac{\Omega_{0m,u}-\Omega_{0m,d}}{\Omega_{0m,u}+\Omega_{0m,d}}$)\ &  \cite{Xiaofeng},\cite{a128}\\
\hline %inserts single line
7  &hemisphere comparison method&$\frac{\Delta q_{0}}{\bar{q_{0}}}=
2(\frac{q_{0,u}-q_{0,d}}{q_{0,u}+q_{0,d}}$)&  \cite{Cai0} \\
\hline %inserts single line
8  & luminosity-distance
&$d_{L}(z,\theta)=\frac{1+z}{H0}\int_{A(z)}^{1}\frac{dA}{A^{2}\bar{H}}\frac{(1-e^{2})^{1/6}}{(1-e^{2}cos\theta)^{1/2}}$&  \cite{a122}\\
9 &in ellipsoidal
universe& $1+z=\frac{1}{A}\frac{(1-e^{2}sin\theta)^{1/2}}{(1-e^{2})^{1/3}}$&\\
\hline %inserts single line
10  & $\alpha$(dipole+monopole) fit approach&$\frac{\Delta\alpha}{\alpha}=A \cos\theta +B$ &  \cite{Antoniou},\cite{Webb1} \\
\hline %inserts single line
11  &measured (perturbed) luminosity-distance&$D_{L}=(1+2\hat{n}.\vec{v_{s}})D_{0L}$ ,$v_{s}$=peculiar velocities &  \cite{De-Chang} \\
\hline %inserts single line
12  &dipole fit approach&$\frac{d_{L}(z)-d^{0}_{L}(z)}{d^{0}_{L}(z)}=g(z)(\hat{z}.\hat{n}$) &  \cite{Cai0} \\
\hline %inserts single line

\end{tabular}\\
  % is used to refer this table in the text
\end{table}\\
A potpourri list of pertinent studies is provided in Table I.
  Ref. \cite{Cai0},  have  taken the deceleration parameter $q_{0}$ as the diagnostic to quantify the
anisotropy level in the $\omega CDM$ model.
The authors of Ref.\cite{a128} constructed a direction-dependent
dark energy model based on the isotropic background described
by the $\Lambda CDM ,\omega CDM $ and $CPL$ models and employed  the Union2 dataset to constrain the anisotropy direction and strength of modulation. They found the best-fitting value of the maximum deviation
direction from the isotropic background, which was not sensitive to the details of isotropic dark
energy models.
 Ref.\cite{Wang} have studied  dipolar anisotropic expansion with cosmographic parameters, and found the preffered direction of $(l=309^{\circ},b=-8.6^{\circ})$.
 The athours of \cite{Xiaofeng}  choosed two simple cosmological
models, $ \Lambda CDM $ and $ \omega CDM$ for the hemisphere comparison
approach, and $\Lambda CDM $ for the dipole fit. In the first approach,
they used the matter density and the equation of state of dark
energy as the diagnostic qualities in the $\Lambda CDM$ and $ \omega CDM$,
respectively. In the second method, they employed  distance modulus as the diagnostic quality in $\Lambda CDM$, and  found the preferred
direction  of $(l=307^{\circ},b=-14^{\circ})$.\\

 Here, we investigate the anisotropy expansion of the universe using the model-independent cosmography method. The cosmography method brings forward a part of cosmology
which does not postulate any precedential cosmological model.
Thus, it can be thought as a model-independent way
to fix the constraints on the universe's dynamics at late times
through the use of a set of parameters; namely the cosmographic
set (CS)\cite{Capozziello}-\cite{Cattoen}.
The outline of the paper is as follows. The cosmographic apparatus is reviewed in Section. II. The luminosity distance redshift relation in the anisotropic Universe will be extracted in Section III.   The  acquisition way of the matrix $M(n\times n)$  is introduced to convert the Taylor series coefficients of isotropy  luminosity distance  to   anisotropy factors of luminosity distance. The numerical results are given in Section. IV and in order to discuss and compare with other works. Arguments are
given in section V.

\section{The cosmographic apparatus}
One of the basic relations in modern cosmology and cosmography is the luminosity distance  redshift $d_{L}(z)$ used in the definition of redshift in an isotropic Universe
 \begin{equation}\label{is}
1+z=\frac{a(t_{0})}{a(t)}
\end{equation}
The following relation is also retained  from the cosmographic approach \cite{Capozziello}-\cite{Cattoen};
\begin{eqnarray}\label{iso}
d_{L}(z) = \frac{c z}{H_{0}} \left\{ \mathcal{D}_{L}^{0} +
\mathcal{D}_{L}^{1} \ z + \mathcal{D}_{L}^{2} \ z^{2} +
\mathcal{D}_{L}^{3} \ z^{3} + \mathcal{D}_{L}^{4} \ z^{4} +
\emph{O}(z^{5}) \right\}
\end{eqnarray}
In which; {\setlength\arraycolsep{0.2pt}
\begin{eqnarray}
\mathcal{D}_{L}^{0} &=& 1 \\
\mathcal{D}_{L}^{1} &=& - \frac{1}{2} \left(-1 + q_{0}\right) \\
\mathcal{D}_{L}^{2} &=& - \frac{1}{6} \left(1 - q_{0} - 3q_{0}^{2} + j_{0} + \frac{k c^{2}}{H_{0}^{2}a_{0}^{2}}\right) \\
\mathcal{D}_{L}^{3} &=& \frac{1}{24} \left(2 - 2 q_{0} - 15
q_{0}^{2} - 15 q_{0}^{3} + 5 j_{0} + 10 q_{0} j_{0} + s_{0} +
\frac{2 k c^{2} (1 + 3 q_{0})}{H_{0}^{2} a_{0}^{2}}\right)\\
\mathcal{D}_{L}^{4} &=& \frac{1}{120} \left[ -6 + 6 q_{0} + 81
q_{0}^{2} + 165 q_{0}^{3} + 105 q_{0}^{4} - 110 q_{0} j_{0} - 105
q_{0}^{2} j_{0} - 15 q_{0} s_{0} + \right. \\
&-& \left.  27 j_{0} + 10 j_{0}^{2} - 11 s_{0} - l_{0} -
\frac{5kc^{2}(1 + 8 q_{0} + 9 q_{0}^{2} - 2 j_{0})}{a_{0}^{2}
H_{0}^{2}}\right]\nonumber
\end{eqnarray}}
{\setlength\arraycolsep{0.2pt}
Where the cosmographic  parameters are defined as;

\begin{eqnarray}
H(t) \equiv + \frac{1}{a}\frac{da}{dt}\, ,\ \
q(t) \equiv - \frac{1}{a}\frac{d^{2}a}{dt^{2}}\frac{1}{H^{2}}\,\ \
,\ \
j(t) \equiv + \frac{1}{a}\frac{d^{3}a}{dt^{3}}\frac{1}{H^{3}}\,
,
\ \
s(t) \equiv + \frac{1}{a}\frac{d^{4}a}{dt^{4}}\frac{1}{H^{4}}\,
,\ \
l(t)\equiv + \frac{1}{a}\frac{d^{5}a}{dt^{5}}\frac{1}{H^{5}}\,
,
\end{eqnarray}}

\section{The luminosity distance redshift relation in the anisotropic Universe}
In the anisotropic cosmological model, the redshift of an object located in the $\hat{\textbf{p}}$
direction  of each SNe Ia sample to galactic coordinates and at time t can be modified as
\begin{equation}\label{aniso}
1+\tilde{z}(t,\hat{\textbf{p}})=\frac{a(t_{0})}{a(t)}(1-A(\hat{\textbf{n}}.\hat{\textbf{p}}))
\end{equation}
where $| A |<< 1$ represents the magnitude of anisotropy of the universe and $\hat{\textbf{n}}$ is the
direction of the privileged axis

In the situation, by succession of    $ \tilde{z}$ rather than $z$ in equation (\ref{iso}), the anisotropy luminosity distance will be equal to,
\begin{eqnarray}\label{10}
d_{L}(\tilde{z}) = \frac{c \tilde{z}}{H_{0}} \left\{ \mathcal{D}_{L}^{0} +
\mathcal{D}_{L}^{1} \ \tilde{z} + \mathcal{D}_{L}^{2} \ \tilde{z}^{2} +
\mathcal{D}_{L}^{3} \ \tilde{z}^{3} + \mathcal{D}_{L}^{4} \ \tilde{z}^{4} +
\emph{O}(\tilde{z}^{5}) \right\}
\end{eqnarray}
Here we suppose that $H_{0}$ is constant. However angular anisotropy of $H_{0}$ has been  studied  in several works (see for example\cite{1}).
Using equations (\ref{aniso}) and (\ref{is}), and since $A$ is a very small magnitude,
\begin{equation}\label{modify}
1+z=\frac{1 + \tilde{z}}{1-A(\hat{\textbf{n}}.\hat{\textbf{p}})}=(1 + \tilde{z})(1+A(\hat{\textbf{n}}.\hat{\textbf{p}})),
\end{equation}
where for $A\ll1$ , we have used $(\frac{1}{1-x}\simeq1+x) $.
Therefore, the equation takes the modified redshift $\tilde{z}$ in the following form,
\begin{equation}\label{dd}
\tilde{z}\simeq\frac{1 + z}{1+A(\hat{\textbf{n}}.\hat{\textbf{p}})}-1\simeq
\frac{z-A(\hat{\textbf{n}}.\hat{\textbf{p}})}{1+A(\hat{\textbf{n}}.\hat{\textbf{p}})}\simeq
\frac{z}{1+A(\hat{\textbf{n}}.\hat{\textbf{p}})}-\frac{A(\hat{\textbf{n}}.\hat{\textbf{p}})}
{1+A(\hat{\textbf{n}}.\hat{\textbf{p}})}\simeq z-\frac{A(\hat{\textbf{n}}.\hat{\textbf{p}})}
{1+A(\hat{\textbf{n}}.\hat{\textbf{p}})},
\end{equation}
where for very small value of $A$ ,we have used( $1+A(\hat{\textbf{n}}.\hat{\textbf{p}})\simeq1$) and consequently ( $\frac{z}{1+A(\hat{\textbf{n}}.\hat{\textbf{p}})}\simeq z$).
By considering $\delta=\frac{A(\hat{\textbf{n}}.\hat{\textbf{p}})}{1+A(\hat{\textbf{n}}.\hat{\textbf{p}})}$ and substituting $(\tilde{z}=z-\delta)$ into equation (\ref{10}),
we can rewrite this equation as
\begin{eqnarray}
d_{L}(\tilde{z}) = \frac{c( z-\delta)}{H_{0}} \left\{ \mathcal{D}_{L}^{0} +
\mathcal{D}_{L}^{1} \ ( z-\delta) + \mathcal{D}_{L}^{2} \ ( z-\delta)^{2} +
\mathcal{D}_{L}^{3} \ ( z-\delta)^{3} + \mathcal{D}_{L}^{4} \ ( z-\delta)^{4} +......
\mathcal{D}_{L}^{n}( z-\delta)^{n} \right\}
\end{eqnarray}
\emph{By expanding $(z-\delta)^{m}$ using binomial theorem,}

\begin{eqnarray}\label{dlin3}
(x+y)^{m}=\sum_{k=0}^{m} \left(
      \begin{array}{c}
        m \\
        k \\
      \end{array}
    \right)x^{m-k}y^{k}
\end{eqnarray}
The luminosity distance $d_{L}$ will be obtained in terms of $z$ with new constant coefficients $\mathcal{\acute{D}}_{L}^{m}$ as

\begin{eqnarray}\label{dlin}
d_{L}(\tilde{z}) = \frac{c z}{H_{0}} \left\{ \mathcal{\acute{D}}_{L}^{0} +
\mathcal{\acute{D}}_{L}^{1} \ z + \mathcal{\acute{D}}_{L}^{2} \ z^{2} +
\mathcal{\acute{D}}_{L}^{3} \ z^{3} + \mathcal{\acute{D}}_{L}^{4} \ z^{4} +...
\mathcal{\acute{D}}_{L}^{n} \ z^{n}) \right\}
\end{eqnarray}

  Where, the anisotropic coefficients can be obtain by expanding the luminosity distance to the nth order of redshift $\tilde{z}$,  from the corresponding isotropic factors through the  transformation matrix $M$ as,
\begin{center}
$\left(
  \begin{array}{c}
    \mathcal{\acute{D}}_{L}^{n} \\
    \mathcal{\acute{D}}_{L}^{n-1}\\
    \mathcal{\acute{D}}_{L}^{n-2} \\
    . \\
    . \\
     \mathcal{\acute{D}}_{L}^{0} \\
  \end{array}
\right)$=M$\left(
  \begin{array}{c}
    \mathcal{D}_{L}^{n} \\
    \mathcal{D}_{L}^{n-1}\\
    \mathcal{D}_{L}^{n-2}\\
    . \\
    . \\
    \mathcal{D}_{L}^{0} \\
  \end{array}
\right)$\\
\end{center}

So that, the matrix $M$ would be,\\
\\

$\left(
  \begin{array}{cccccc}
    \left(
      \begin{array}{c}
        n+1 \\
        0 \\
      \end{array}
    \right)
     & 0 & 0 & 0 & 0& 0 \\
    \left(
      \begin{array}{c}
        n+1 \\
        1 \\
      \end{array}
    \right)\delta & -\left(
      \begin{array}{c}
        n \\
        0 \\
      \end{array}
    \right) & 0 & 0 & 0& 0 \\
    \left(
      \begin{array}{c}
        n+1 \\
        2 \\
      \end{array}
    \right)\delta^{2} &- \left(
      \begin{array}{c}
        n \\
        1\\
      \end{array}
    \right)\delta & \left(
      \begin{array}{c}
        n-1 \\
        0 \\
      \end{array}
    \right) & 0 & 0 & 0\\
    \left(
      \begin{array}{c}
        n+1 \\
        3 \\
      \end{array}
    \right)\delta^{3} & -\left(
      \begin{array}{c}
        n \\
        2 \\
      \end{array}
    \right)\delta^{2} & \left(
      \begin{array}{c}
        n-1 \\
        1 \\
      \end{array}
    \right)\delta & -\left(
      \begin{array}{c}
        n-2 \\
        0 \\
      \end{array}
    \right) & 0 & 0\\
    . & . & . & . & . & 0\\
    \left(
      \begin{array}{c}
        n+1 \\
        n \\
      \end{array}
    \right)\delta^{n} & -\left(
      \begin{array}{c}
        n \\
        n-1 \\
      \end{array}
    \right)\delta^{n-1} & \left(
      \begin{array}{c}
        n-1 \\
        n-2 \\
      \end{array}
    \right)\delta^{n-2} & -\left(
      \begin{array}{c}
        n-2 \\
        n-3 \\
      \end{array}
    \right)\delta^{n-3} & .& (-1)^{j+1}\left(
      \begin{array}{c}
        n+1-n \\
        n-n \\
      \end{array}
    \right)\delta^{0} \\
  \end{array}
\right)$\\
\\
Neglecting higher order terms of $\delta^{n}$ with $n > 1$,
\begin{eqnarray}
\mathcal{\acute{D}}_{L}^{0} &=& (\mathcal{D}_{L}^{0}-2\mathcal{D}_{L}^{1}\delta) \\
\mathcal{\acute{D}}_{L}^{1} &=& (\mathcal{D}_{L}^{1}-3\mathcal{D}_{L}^{2}\delta \\
\mathcal{\acute{D}}_{L}^{2} &=& (\mathcal{D}_{L}^{2}-4\mathcal{D}_{L}^{3}\delta) \\
\mathcal{\acute{D}}_{L}^{3} &=& (\mathcal{D}_{L}^{3}-5\mathcal{D}_{L}^{4}\delta)\\
. &=& .\\
. &=& .\\
\mathcal{\acute{D}}_{L}^{n-1} &=&(\mathcal{D}_{L}^{n-1}-(n+1)\mathcal{D}_{L}^{n+1}\delta)\\
\mathcal{\acute{D}}_{L}^{n} &=& \mathcal{D}_{L}^{n}
\end{eqnarray}
 For example, for series expansion of $d_{L}(\tilde{z})$ to the 5th order in redshift $\tilde{z}$,
we have
\begin{eqnarray}\label{dlin}
d_{L}(\tilde{z}) = \frac{c z}{H_{0}} \left\{ \mathcal{\acute{D}}_{L}^{0} +
\mathcal{\acute{D}}_{L}^{1} \ z + \mathcal{\acute{D}}_{L}^{2} \ z^{2} +
\mathcal{\acute{D}}_{L}^{3} \ z^{3} + \mathcal{\acute{D}}_{L}^{4} \ z^{4} +
\emph{O}(z^{5}) \right\}
\end{eqnarray}
where
\begin{eqnarray}
\mathcal{\acute{D}}_{L}^{0} &=& (\mathcal{D}_{L}^{0}-2\mathcal{D}_{L}^{1}\delta+3\mathcal{D}_{L}^{2}\delta^2-4\mathcal{D}_{L}^{3}\delta^3
+5\mathcal{D}_{L}^{4}\delta^4) \\
\mathcal{\acute{D}}_{L}^{1} &=& (\mathcal{D}_{L}^{1}-3\mathcal{D}_{L}^{2}\delta+6\mathcal{D}_{L}^{3}\delta^2-10\mathcal{D}_{L}^{4}\delta^3) \\
\mathcal{\acute{D}}_{L}^{2} &=& (\mathcal{D}_{L}^{2}-4\mathcal{D}_{L}^{3}\delta+10\mathcal{D}_{L}^{4}\delta^2) \\
\mathcal{\acute{D}}_{L}^{3} &=& (\mathcal{D}_{L}^{3}-5\mathcal{D}_{L}^{4}\delta)\\
\mathcal{\acute{D}}_{L}^{4} &=& \mathcal{D}_{L}^{4}
\end{eqnarray}

Accordingly, the theoretical distance modulus
$\mu _{th}(\tilde{z})$ can be  defined as follows,

\begin{eqnarray}\label{distancem}
\mu _{th}(\tilde{z})=5\log_{10} d_{L}(\tilde{z})+42.384-5\log_{10} h_{0}
\end{eqnarray}

\section{Numerical constraints from Union2 data}
The Union2 SnIa dataset \cite{Amanullah} is a compilation consisting of 557 SNe Ia with the redshift range of z =
[0:015; 1:4]. The
angular distribution of the Union 2 dataset in galactic coordinates, has been shown in Fig. 2.
The Union2 data along with directions as presented
in Ref. \cite{Blomqvist} include the SnIa name, the redshift in the
CMB rest frame, the distance modulus and its uncertainties (that involves both of the observational and the intrinsic magnitude scatter).
They also include the equatorial coordinates (right ascension and declination) of each
SnIa. It is straightforward to convert these coordinates
to galactic coordinates or to usual spherical coordinates
$(\theta, \phi) $ in the equatorial or galactic systems\cite{Duffett}. Here, we have converted the equatorial coordinates of each supernova  to the galactic coordinates.  The Cartesian coordinates of the unit vectors $p_{i}$ can also be found corresponding to each quasar with galactic
coordinates $(l, b)$. Since,
\begin{eqnarray}
\hat{p_{i}}= \cos(l_{i})\sin(b_{i})\hat{i}+\sin (l_{i}) \sin( b_{i})\hat{j}+\cos (b_{i})\hat{k}
\end{eqnarray}
The direction
of the privileged axis  can also be  written as,
\begin{eqnarray}
\hat{n}= \cos(l)\sin (b)\hat{i}+\sin (l) \sin( b)\hat{j}+\cos(b)\hat{k}
\end{eqnarray}

In this right, one can use the Union2 compilation  to constrain the
 parameters of model and determine the direction of preferred axis of universe, in the framework of following steps:\\
$\bullet$ {Find the best value  for the cosmographic parameters $\{q_{0},j_{0},s_{0},l_{0},..\}$ in isotropic background to find the best value for expansion's coefficients of isotropic luminosity distance $\{\mathcal{D}_{L}^{0},\mathcal{D}_{L}^{1},\mathcal{D}_{L}^{2}....\mathcal{D}_{L}^{n}\}$.\\
$ \bullet$ Find  the coefficients of anisotropic luminosity distance $\{\mathcal{\acute{D}}_{L}^{0},\mathcal{\acute{D}}_{L}^{1},\mathcal{\acute{D}}_{L}^{2}....\mathcal{\acute{D}}_{L}^{n}\}$  using the transformation matrix $M$.\\
$ \bullet$ Find the anisotropic parameters $(l,b,A)$ and $\delta$ by doing the least $\chi^{2}$  fit to the Union2
data for equation (\ref{distancem}).
\\
In order to  find the best value for the isotropic and anisotropic parameters of the model using SnIa data and examine  the sensitivity of the model with order of luminosity distance expansion,
we perform the $\chi^{2}$  method
for the first, second ,third and fourth order of the expansion.Thus we get the $M_{1}$, $M_{2}$, $M_{3}$ and $M_{4}$ as follows;
\begin{eqnarray}\label{dlinm1}
&M_{1}:d_{L}(\tilde{z}) = \frac{c z}{H_{0}} \left\{ \mathcal{\acute{D}}_{L}^{0} +
\mathcal{\acute{D}}_{L}^{1} \ z   \right\}\\
&M_{2}:d_{L}(\tilde{z}) = \frac{c z}{H_{0}} \left\{ \mathcal{\acute{D}}_{L}^{0} +
\mathcal{\acute{D}}_{L}^{1} \ z + \mathcal{\acute{D}}_{L}^{2} \ z^{2}
  \right\}\\
&M_{3}:d_{L}(\tilde{z}) = \frac{c z}{H_{0}} \left\{ \mathcal{\acute{D}}_{L}^{0} +
\mathcal{\acute{D}}_{L}^{1} \ z + \mathcal{\acute{D}}_{L}^{2} \ z^{2} +
\mathcal{\acute{D}}_{L}^{3} \ z^{3}  \right\}\\
&M_{4}:d_{L}(\tilde{z}) = \frac{c z}{H_{0}} \left\{ \mathcal{\acute{D}}_{L}^{0} +
\mathcal{\acute{D}}_{L}^{1} \ z + \mathcal{\acute{D}}_{L}^{2} \ z^{2} +
\mathcal{\acute{D}}_{L}^{3} \ z^{3} + \mathcal{\acute{D}}_{L}^{4} \ z^{4}
 \right\}\\
\end{eqnarray}
Using the maximum likelihood method (i.e. minimizing),
\begin{eqnarray}
\chi^{2}({h_{0},q_{0},j_{0},...})= \sum_{i=1}^{557}\frac{[\mu^{obs}(z_{i})-\mu^{th}({z_{i}}|{h_{0},q_{0},j_{0},...}]^{2}}{\sigma^{2}(z_{i})}.
\end{eqnarray}
 we can obtain the best value for cosmographic parameters in isotropic background.Table (II) shows these values for in isotropic background in $1-\sigma$ confidence level  for  $M_{1}$ to $M_{4}$ cases.Also  the corresponding likelihood distribution and confidence level for these parameters  have been shown in Figs (1) to (6).\\
It is interesting to note that,  there is a  considerable difference between the best value of cosmographic parameters $\{q_{0},h_{0}\}$ in  the  $M_{1}$ case with higher orders $\{M_{2},M_{3},M_{4}..\}$,however the model is not very sensitive  to order of luminosity distance expansion for higher orders (see figs (1) to (6) and table II).

Using the best value of cosmographic in isotropic background  and  the maximum likelihood method (i.e. minimizing),
\begin{eqnarray}
\chi^{2}(A,l,b)= \sum_{i=1}^{557}\frac{[\mu^{obs}(z_{i})-\mu^{th}({\tilde{z}_{i}}|A ,l,b)]^{2}}{\sigma^{2}(z_{i})}.
\end{eqnarray}
\begin{tabular*}{2.5 cm}{cc}
\includegraphics[scale=.4]{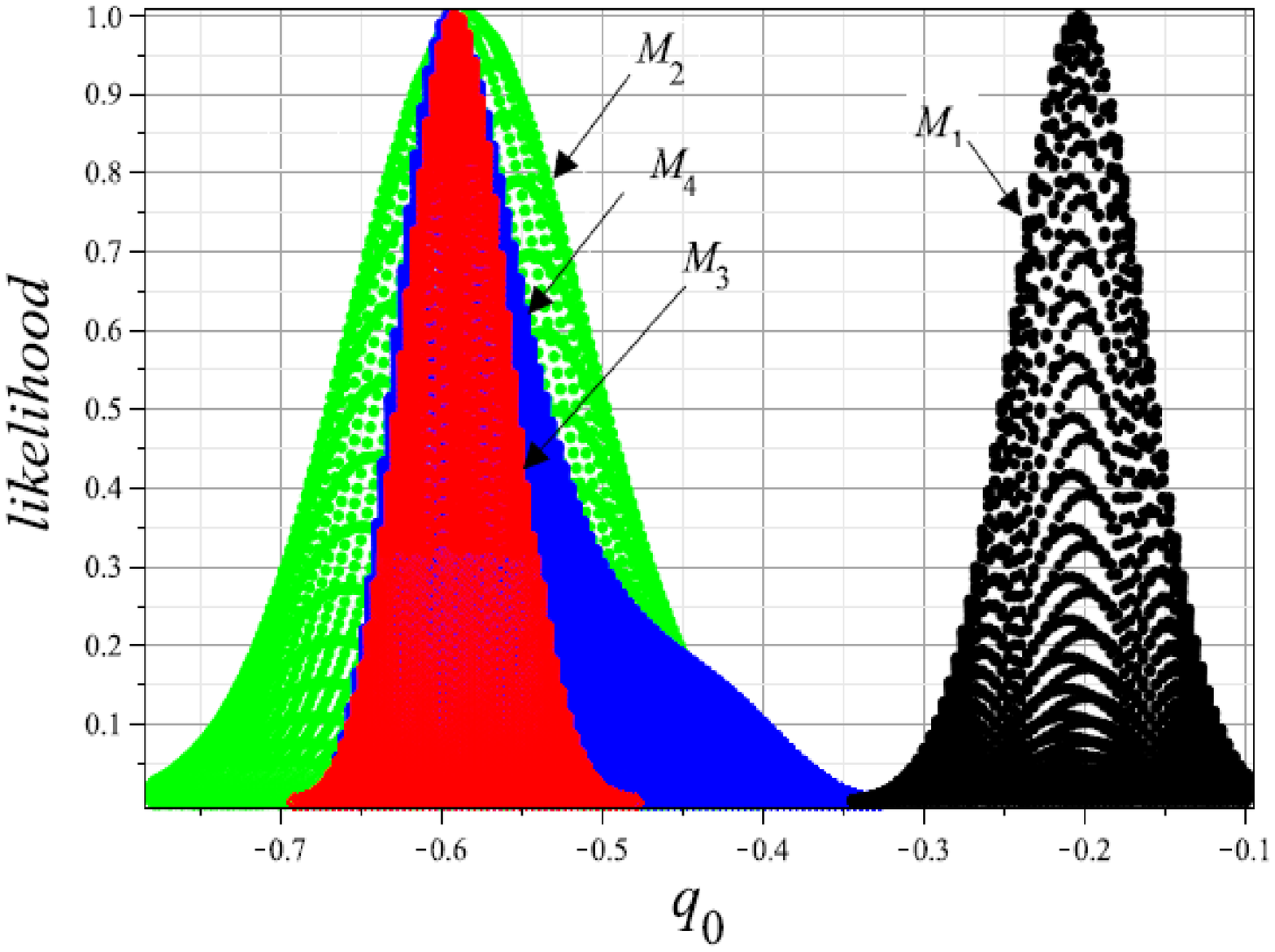}\hspace{0.1 cm}\includegraphics[scale=.4]{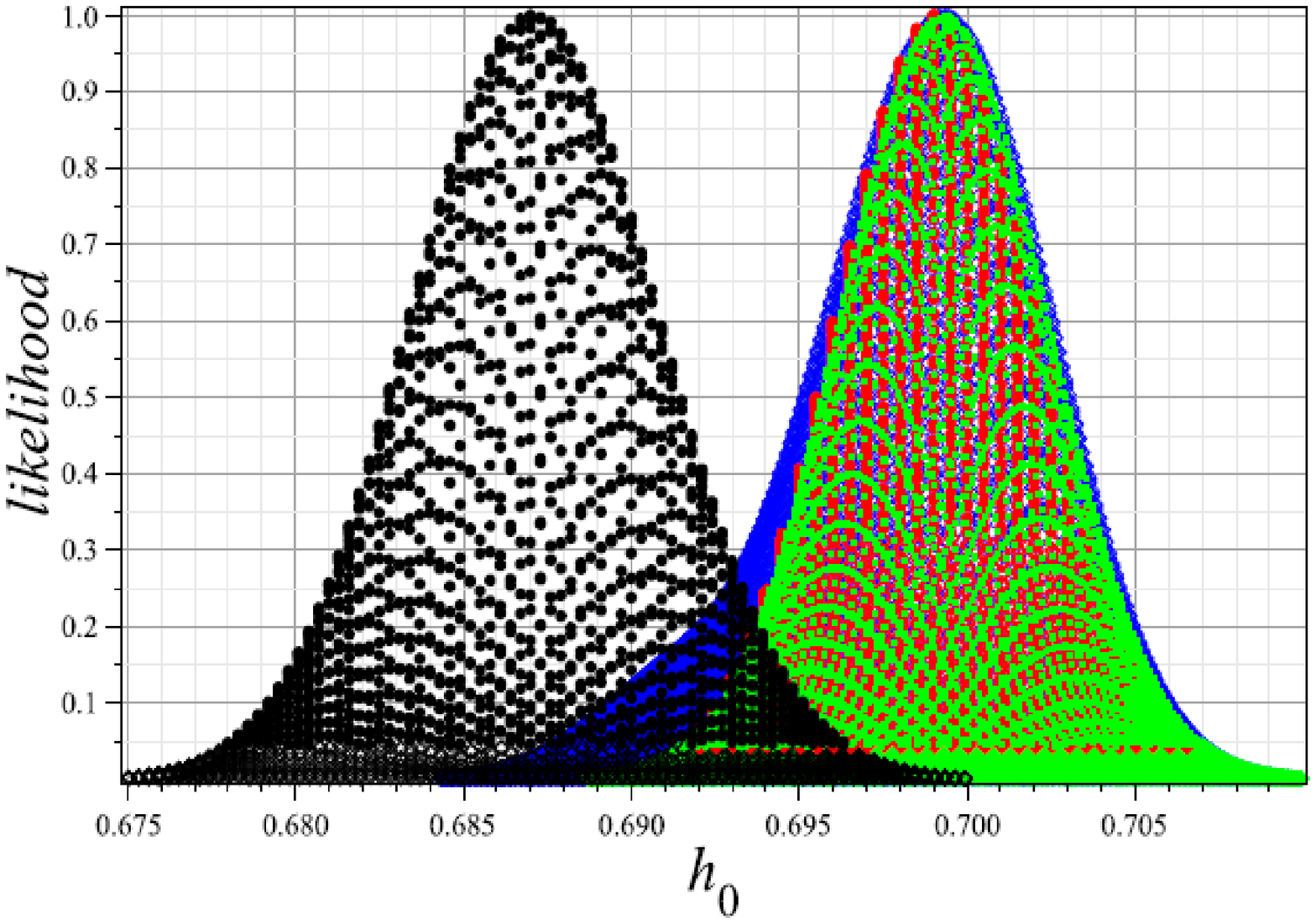}\hspace{0.1 cm} \\
Fig. 1:  The 1-dim likelihood   for parameters $q_{0}$ and $h_{0}$ in $M_{1}$,$M_{2}$,$M_{3}$ and $M_{4}$ cases. \\
The black,green,red and blue colors represent the $M_{1}$,$M_{2}$,$M_{3}$ \\and $M_{4}$ cases respectively \\
\end{tabular*}\\
\begin{tabular*}{2.5 cm}{cc}
\includegraphics[scale=.45]{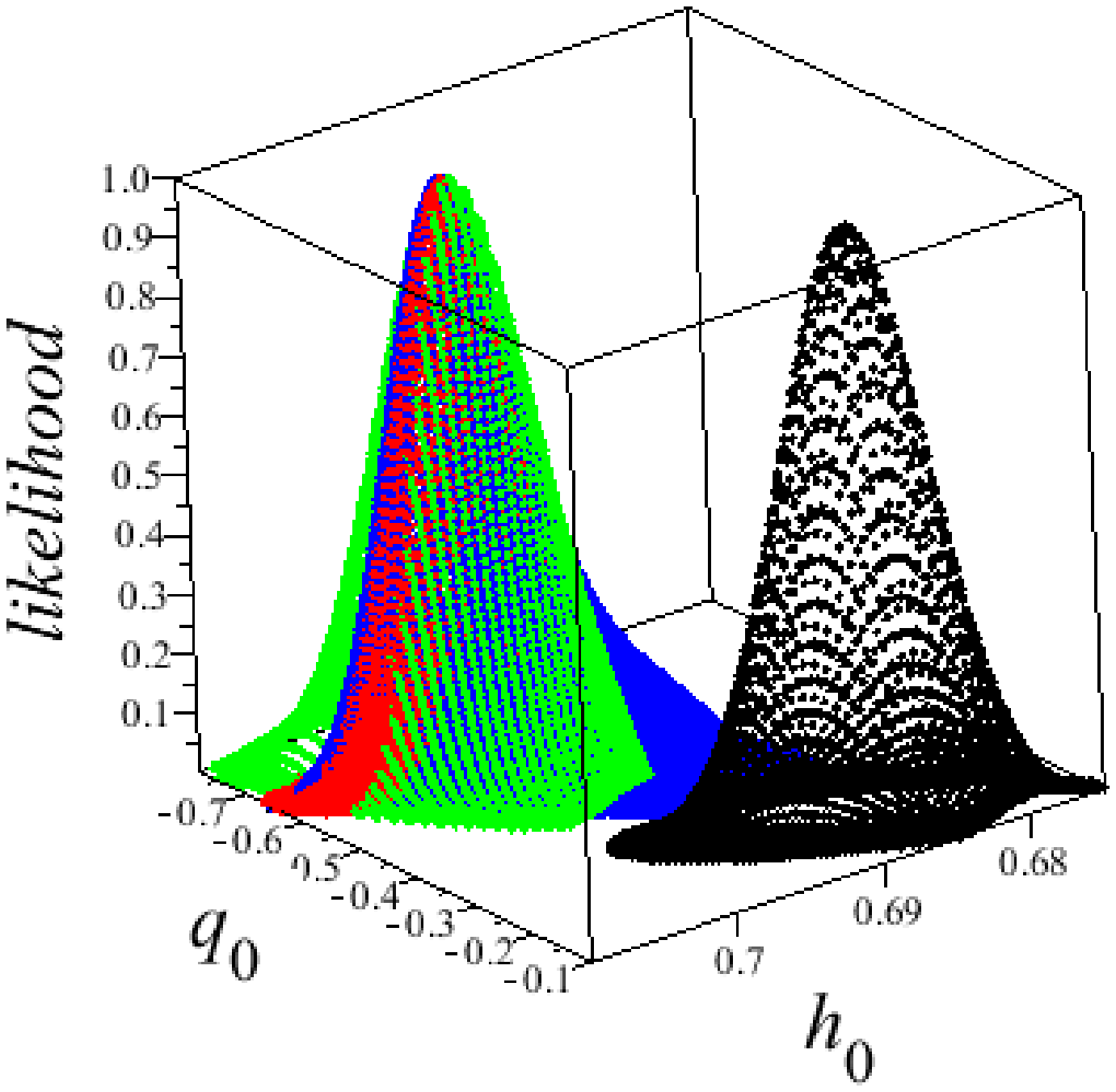}\hspace{0.1 cm}\includegraphics[scale=.4]{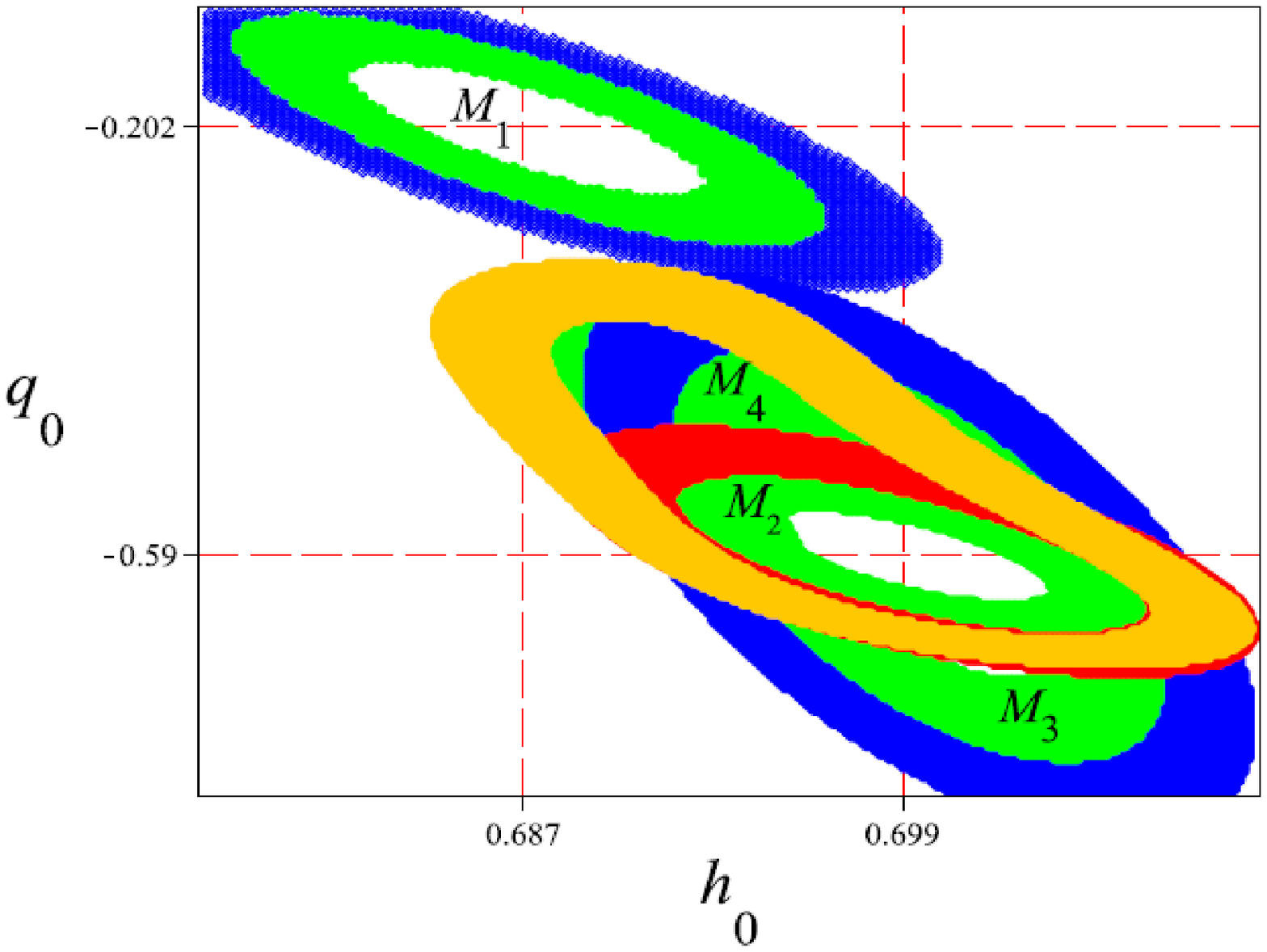}\hspace{0.1 cm} \\
Fig. 2:  The 2-dim likelihood and confidence level for parameters $q_{0}$ and $h_{0}$ \\ in $M_{1}$,$M_{2}$,$M_{3}$ and $M_{4}$ cases. \\
\end{tabular*}\\
\begin{tabular*}{2.5 cm}{cc}
\includegraphics[scale=.4]{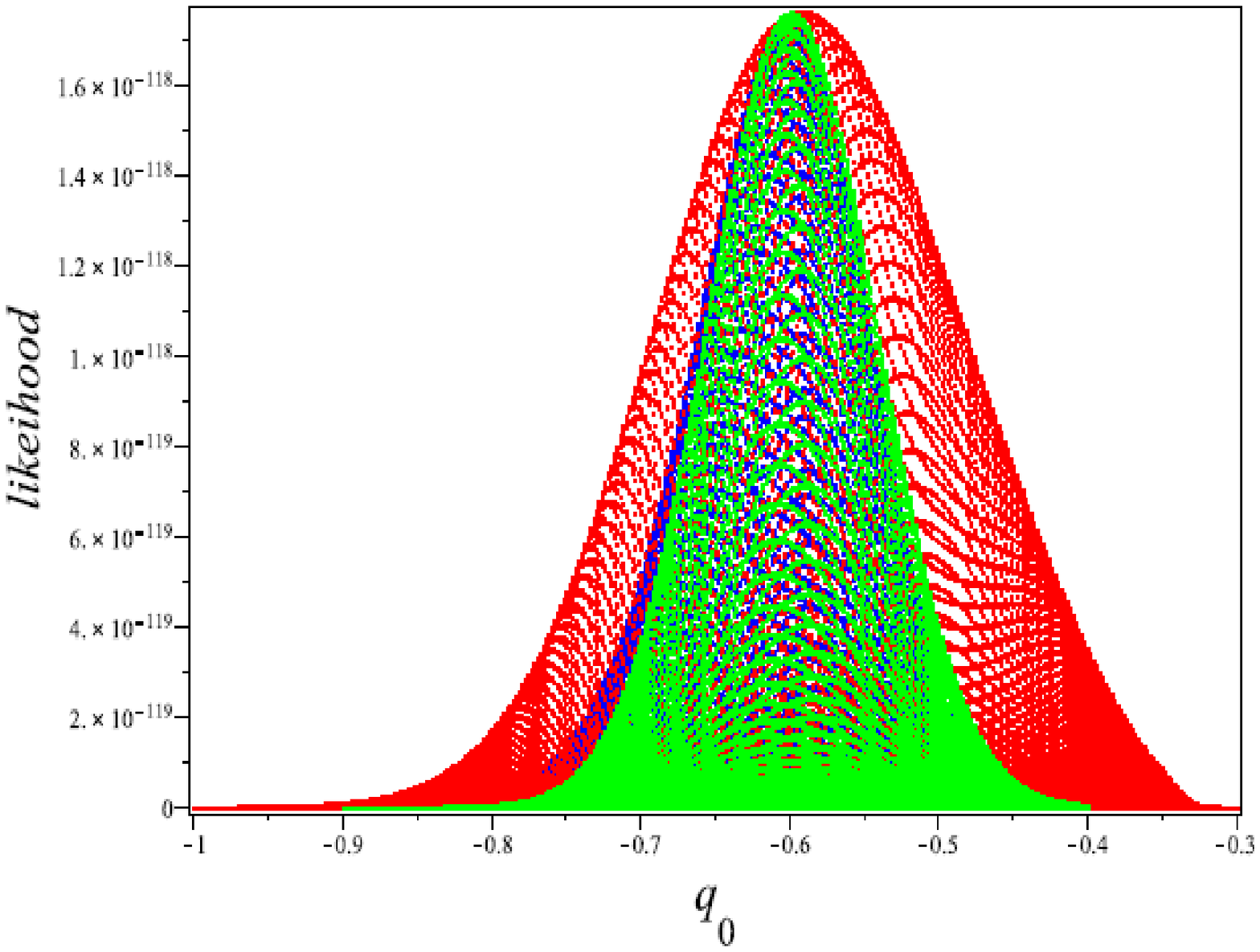}\hspace{0.1 cm}\includegraphics[scale=.4]{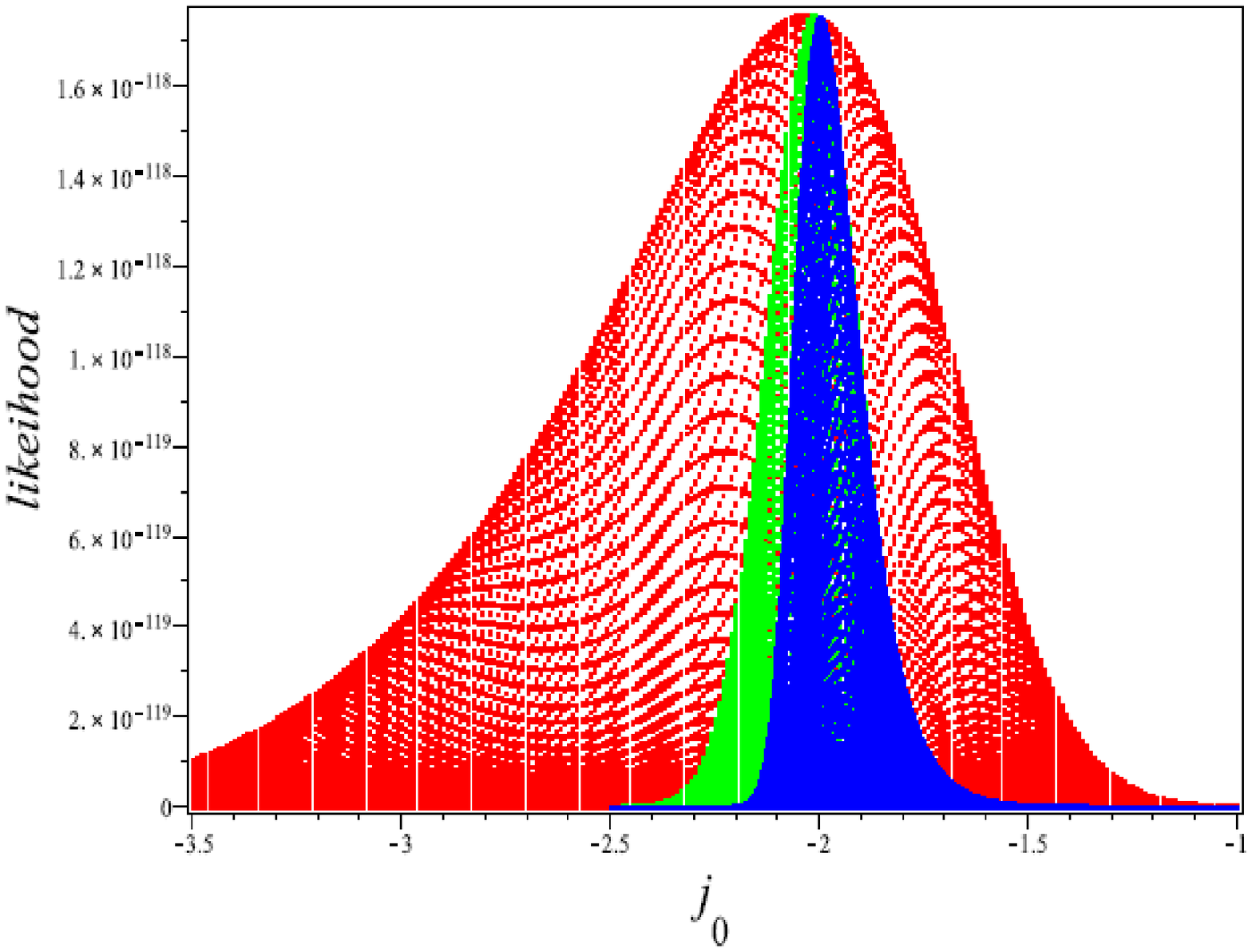}\hspace{0.1 cm} \\
Fig. 3:  The 1-dim likelihood   for parameters $q_{0}$ and $j_{0}$ in $M_{2}$,$M_{3}$ and $M_{4}$ cases.\\
\end{tabular*}\\
\begin{tabular*}{2.5 cm}{cc}
\includegraphics[scale=.5]{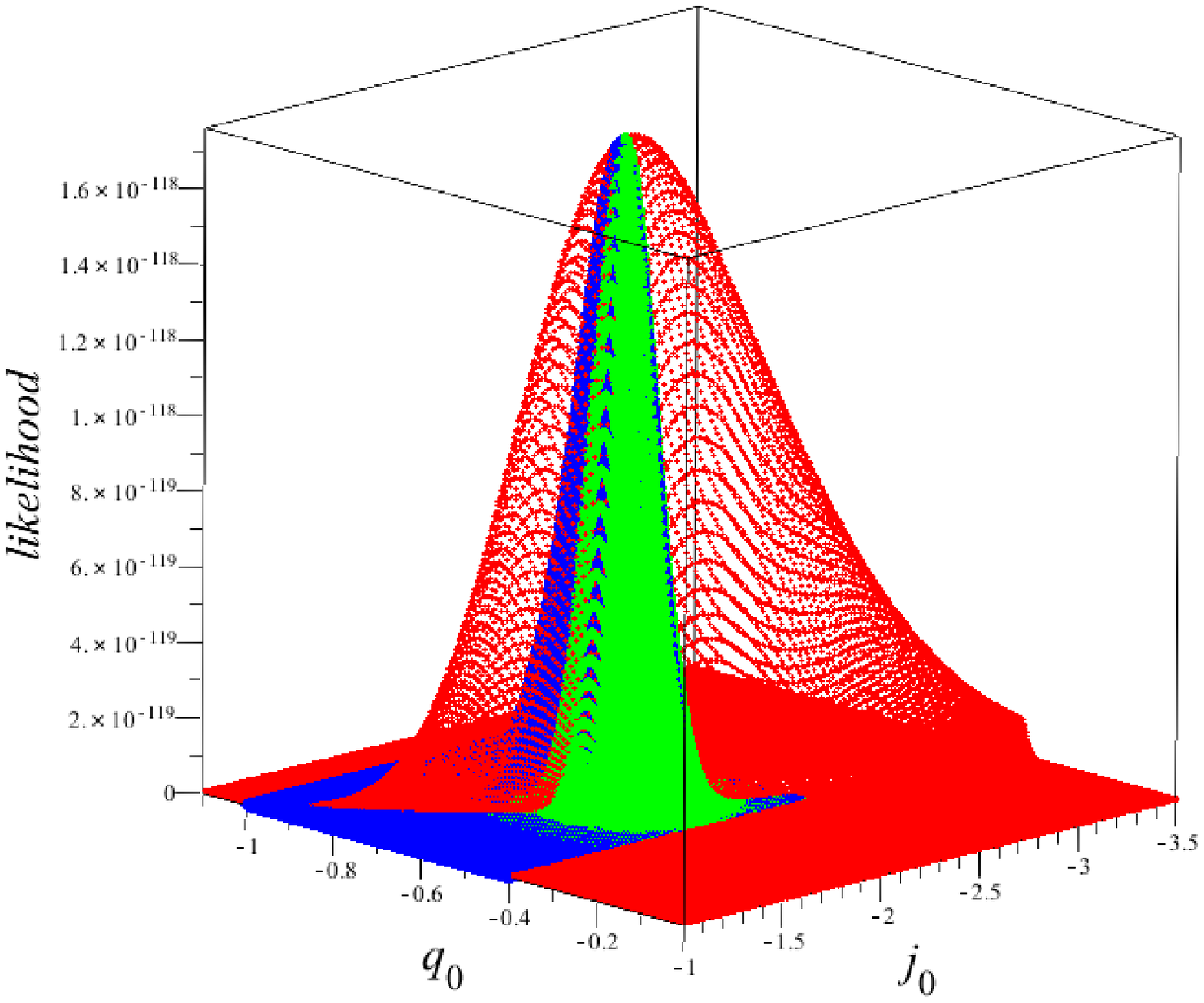}\hspace{0.1 cm} \includegraphics[scale=.4]{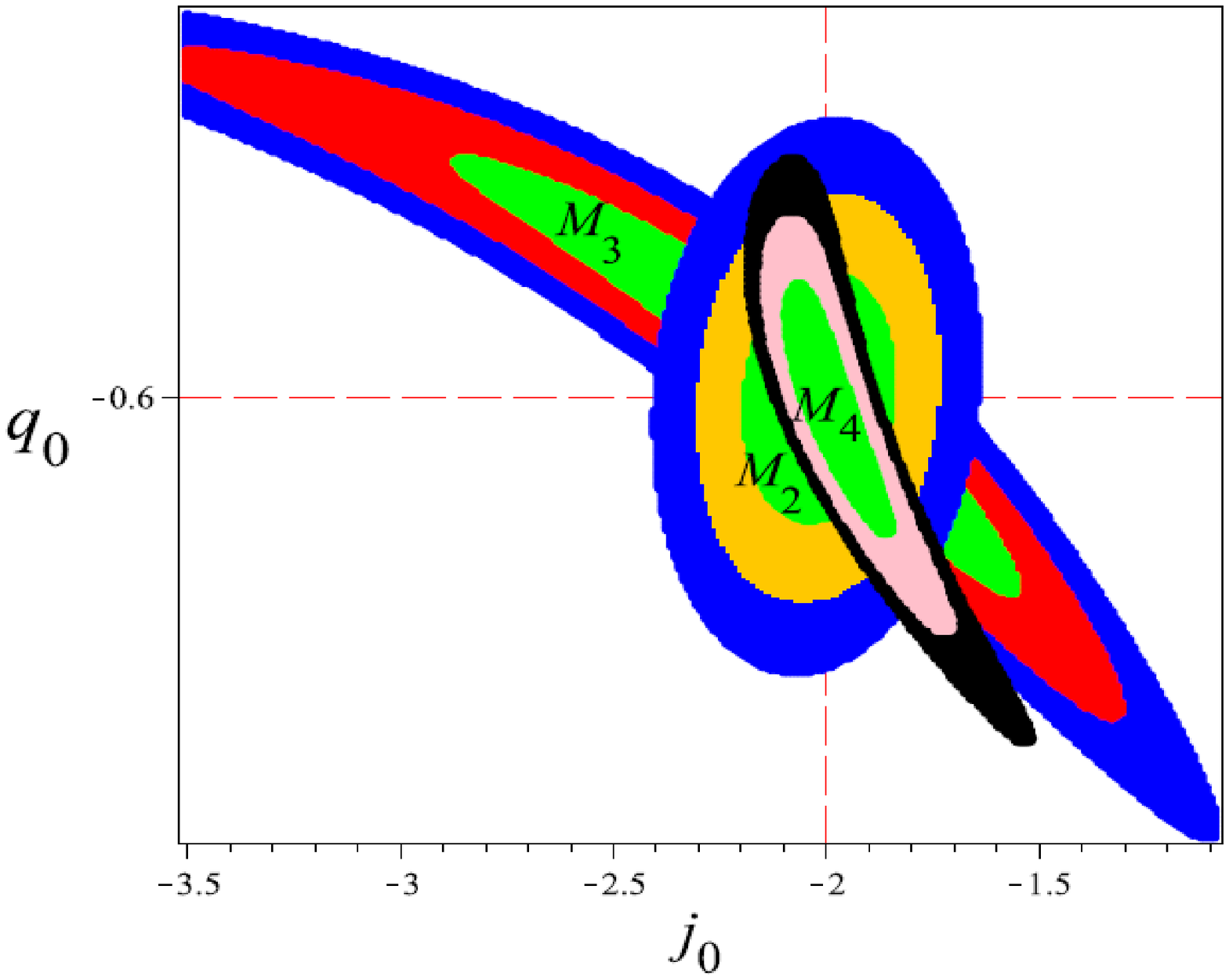}\hspace{0.1 cm} \\
Fig. 4:  The 2-dim  likelihood and confidence level for parameters $q_{0}$ and $j_{0}$ in $M_{2}$,$M_{3}$ and $M_{4}$. \\
\end{tabular*}\\
\begin{tabular*}{2.5 cm}{cc}
\includegraphics[scale=.45]{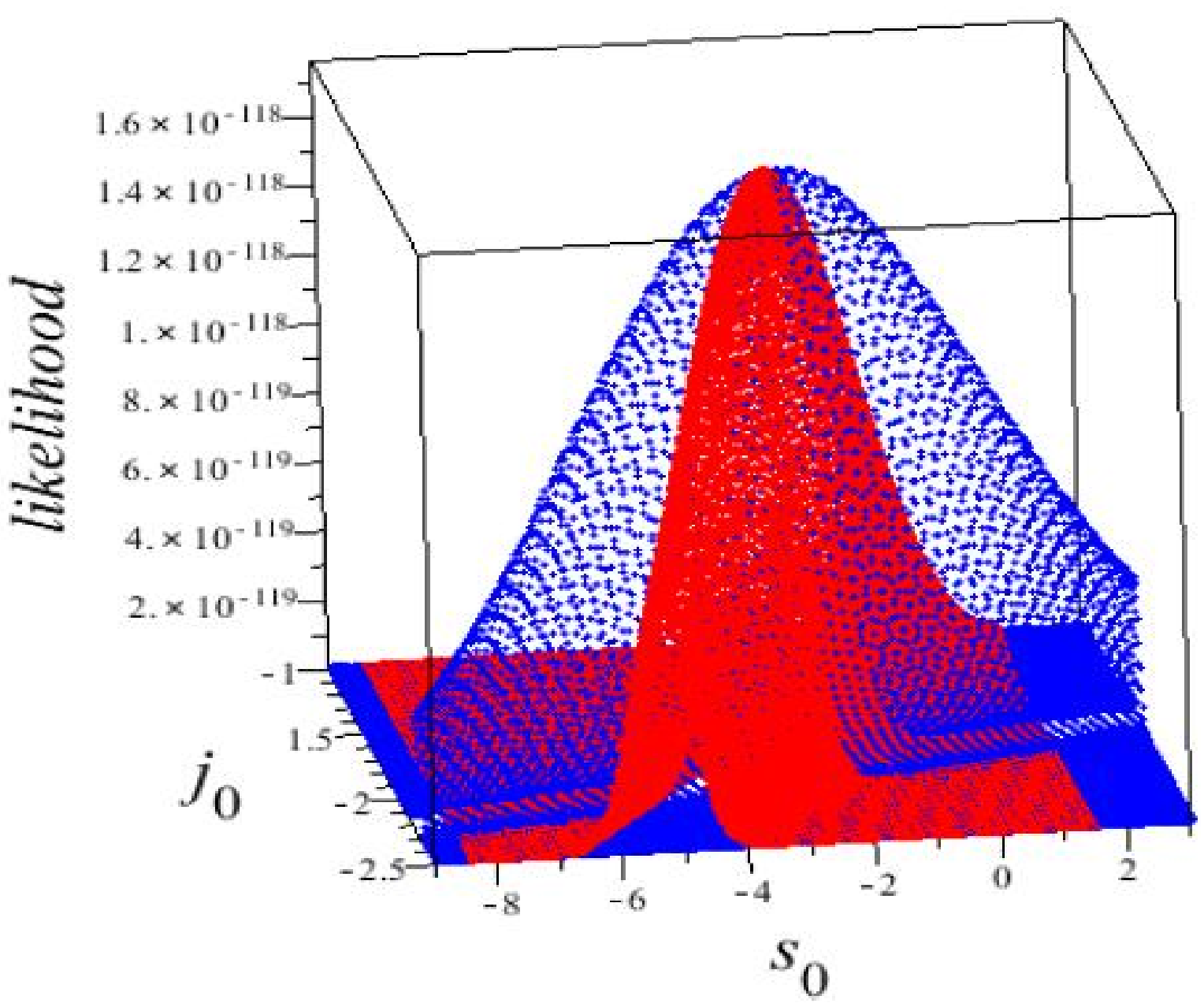}\hspace{0.1 cm} \includegraphics[scale=.45]{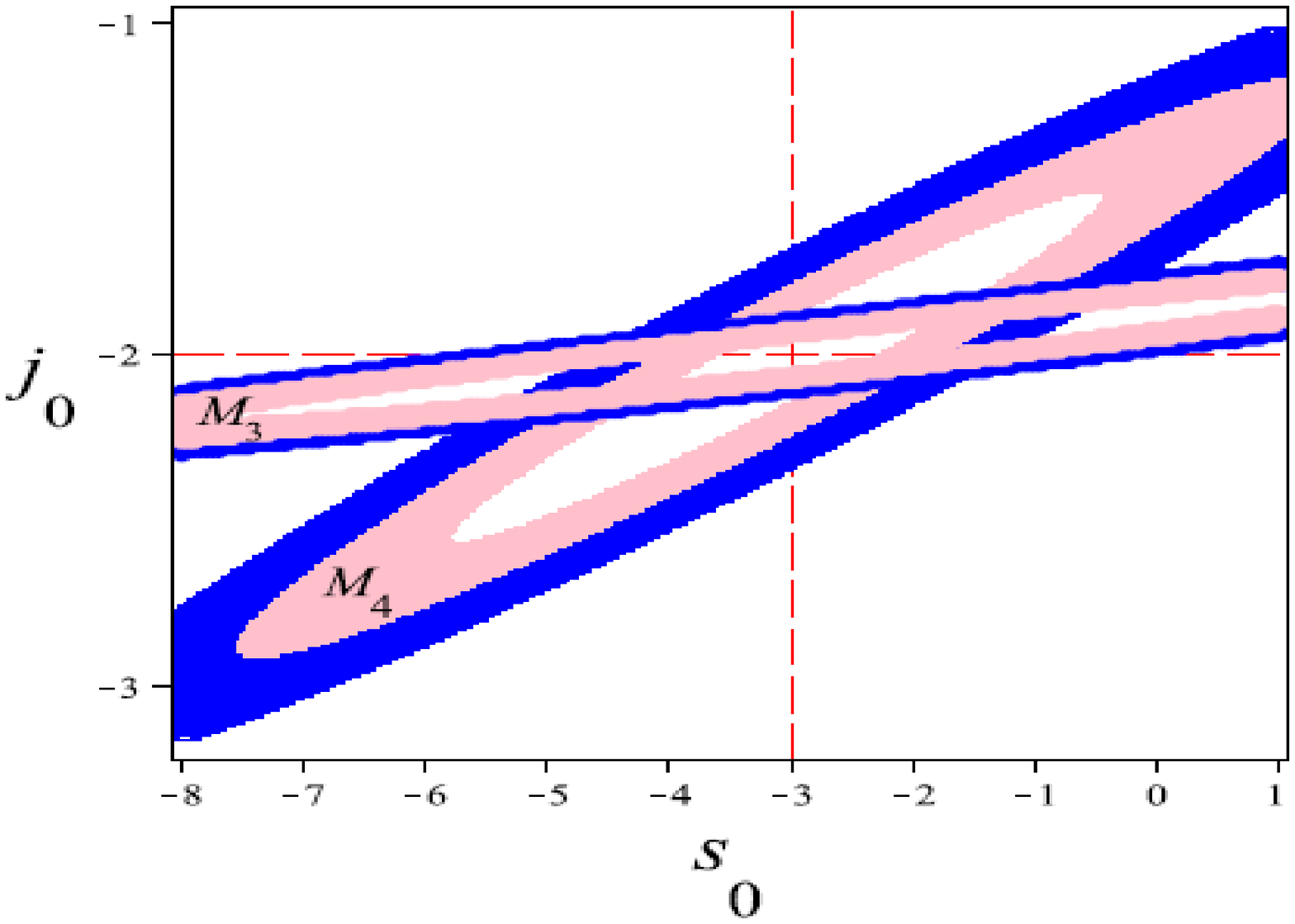}\hspace{0.1 cm} \\
Fig. 5:  The 2-dim  likelihood and confidence level for parameters $s_{0}$ and $j_{0}$   in $M_{2}$,$M_{3}$ and $M_{4}$. \\
\end{tabular*}\\
\begin{tabular*}{2.5 cm}{cc}
\includegraphics[scale=.45]{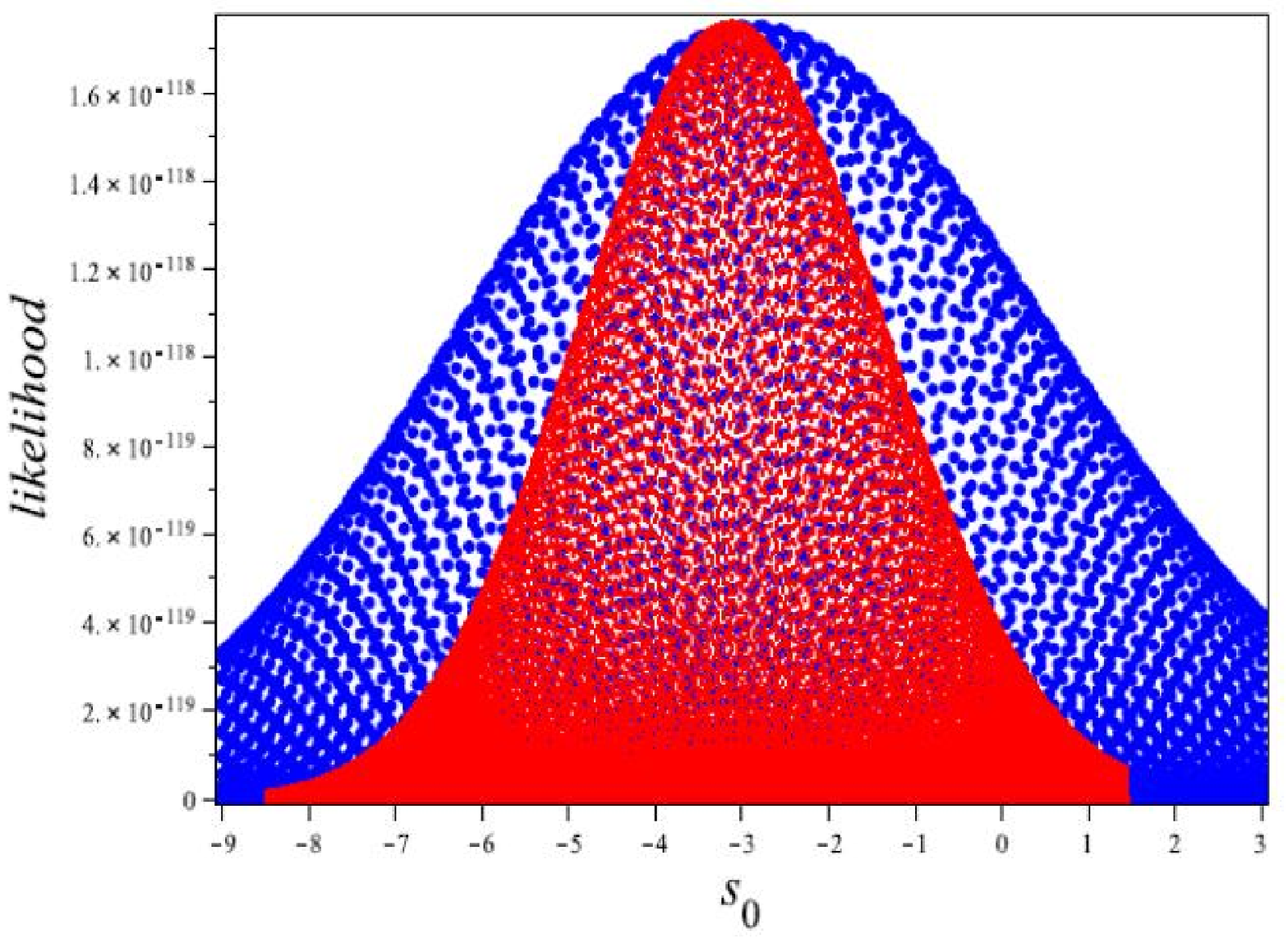}\hspace{0.1 cm} \includegraphics[scale=.42]{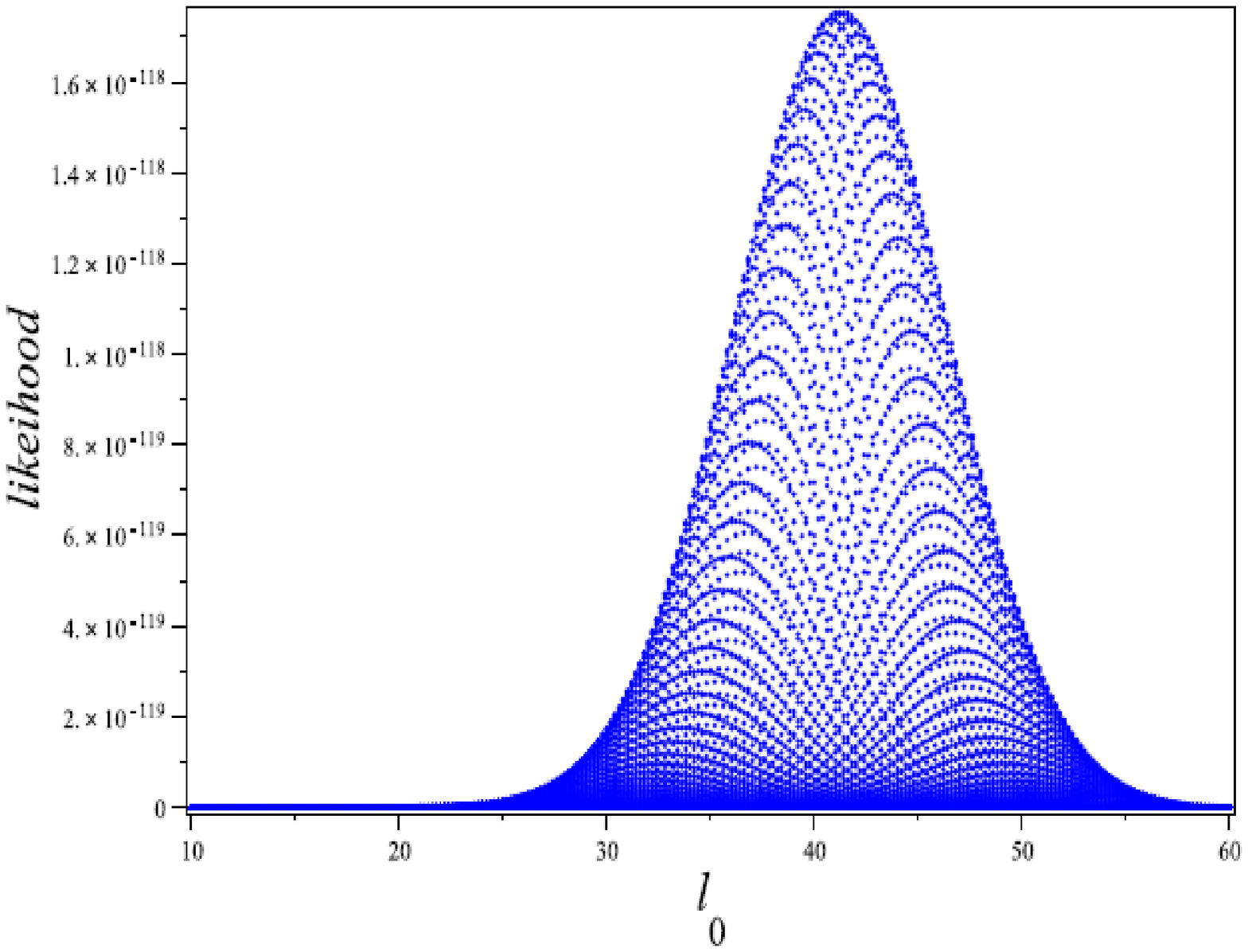}\hspace{0.1 cm} \\
Fig. 6:  The 1-dim  likelihood for parameter $s_{0}$ and $l_{0}$  in $M_{3}$ and $M_{4}$ cases. \\
\end{tabular*}\\

 we can find the magnitude of dipole $A$ and direction of preferred
 axis in galactic coordinates  $(l,b)$.

\begin{table}[ht]
\caption{Best-fitted cosmographic parameters in isotropic background} % % title of Table
\centering % used for centering table
\begin{tabular}{|c| c |c| c |c| c| c| c| c|} % centered columns (6 columns)
\hline %inserts double horizontal lines  \\
\hline %inserts double horizontal lines
  model &  $h_{0}$ &  $q_{0}$  &  $j_{0}$ \ & $s_{0}$\ & $l_{0}$
\ & $\chi^2_{min}$\\ [2ex]
\hline %inserts single line
% inserts table
%heading
\hline % inserts single horizontal line
M1 &$0.687^{+0.0048}_{-0.0048}$  &$-0.204^{+0.10}_{-0.12}$  & $-$ \ & $-$\ & $-$\ &   $559.3851412$ \\
\hline
M2 &$0.699^{+0.0045}_{-0.0045}$  &$-0.582^{+0.06}_{-0.044}$  & $-2.05^{+0.12}_{-0.11}$ \ & $-$\ & $-$\ &   $542.2149085$ \\
\hline %inserts single line
M3 &$0.6993^{+0.0046}_{-0.0046}$ &$-0.594^{+0.08}_{-0.8}$   & $-2.03^{+0.53}_{-0.82}$ \ & $-3^{+5.2}_{-5}$\ & $-$\ &   $542.2794916$ \\%heading
\hline % inserts single horizontal line
M4 &$0.6994^{+0.0017}_{-0.0046}$  &$-.6^{+0.08}_{-0.09}$  & $-2^{+0.1}_{-0.09}$ \ & $-2.9^{+2.9}_{-3.1}$\ & $41.2^{+7.8}_{-8.2}$\ &   $542.2881784$ \\
%heading
\hline % inserts single horizontal line

\hline % inserts single horizontal line
\end{tabular}
\label{table:II} % is used to refer this table in the text
\end{table}\

We construct $2.5\times10^{4}$ such Monte Carlo datasets and
obtain the probability distribution of the dipole
magnitude, as well as the corresponding dipole directions.The distribution of the Union2 SnIa datapoints in galactic coordinates along
with the Dark Energy dipole direction $(l,b)$ in $1-\sigma$ confidence region for different cases $M_{2}$ to $M_{4}$ are shown in Figs.(7) to (9). Interestingly these directions are very close to each other and point towards $(l \simeq297^{\circ}, b = 3^{\circ})$, also the magnitude of the dipole  are found to be $|A|\simeq 10^{-3}$. The likelihood distribution of dipole magnitude has been shown in Fig.(10).
The results of this work are compared by other studies and is approximately consistent with the results of \cite{chang}-\cite{Watkins} and the results of \cite{chang}-\cite{Watkins},  which have been shown in Fig.( 11). \\

\begin{tabular*}{2.5 cm}{cc}
\includegraphics[scale=.8]{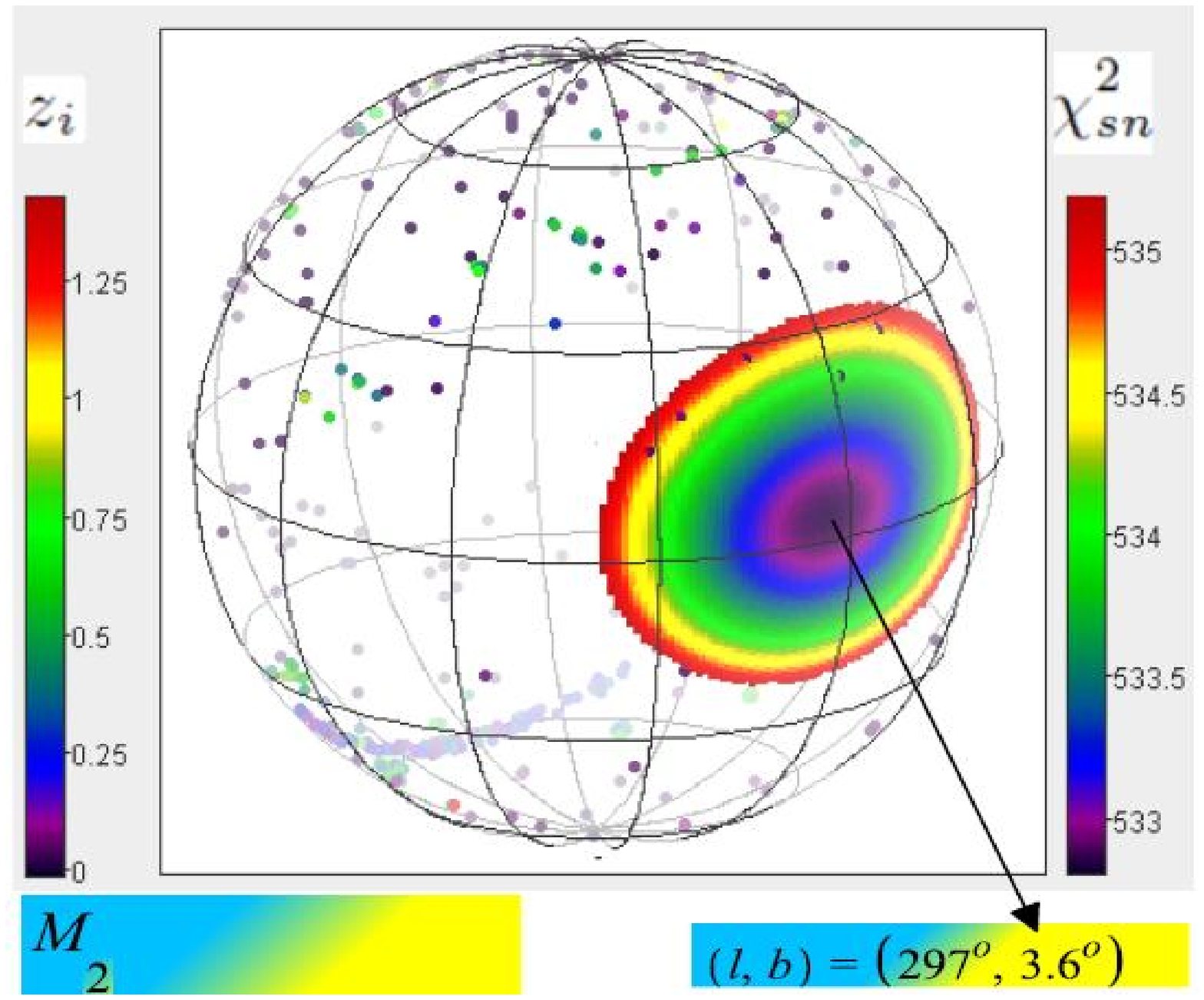}\hspace{0.1 cm}\\
Fig.7: \emph{$1-\sigma$  confidence level for parameters $(l,b)$}. \\ \emph{SNe Ia samples and the dipolar expansion direction in the galactic coordinates} \\ \emph{(used Monte Carlo simulation with} $2.5 \times10^{4}$datasets)
\end{tabular*}\\
\begin{tabular*}{2.5 cm}{cc}
\includegraphics[scale=.8]{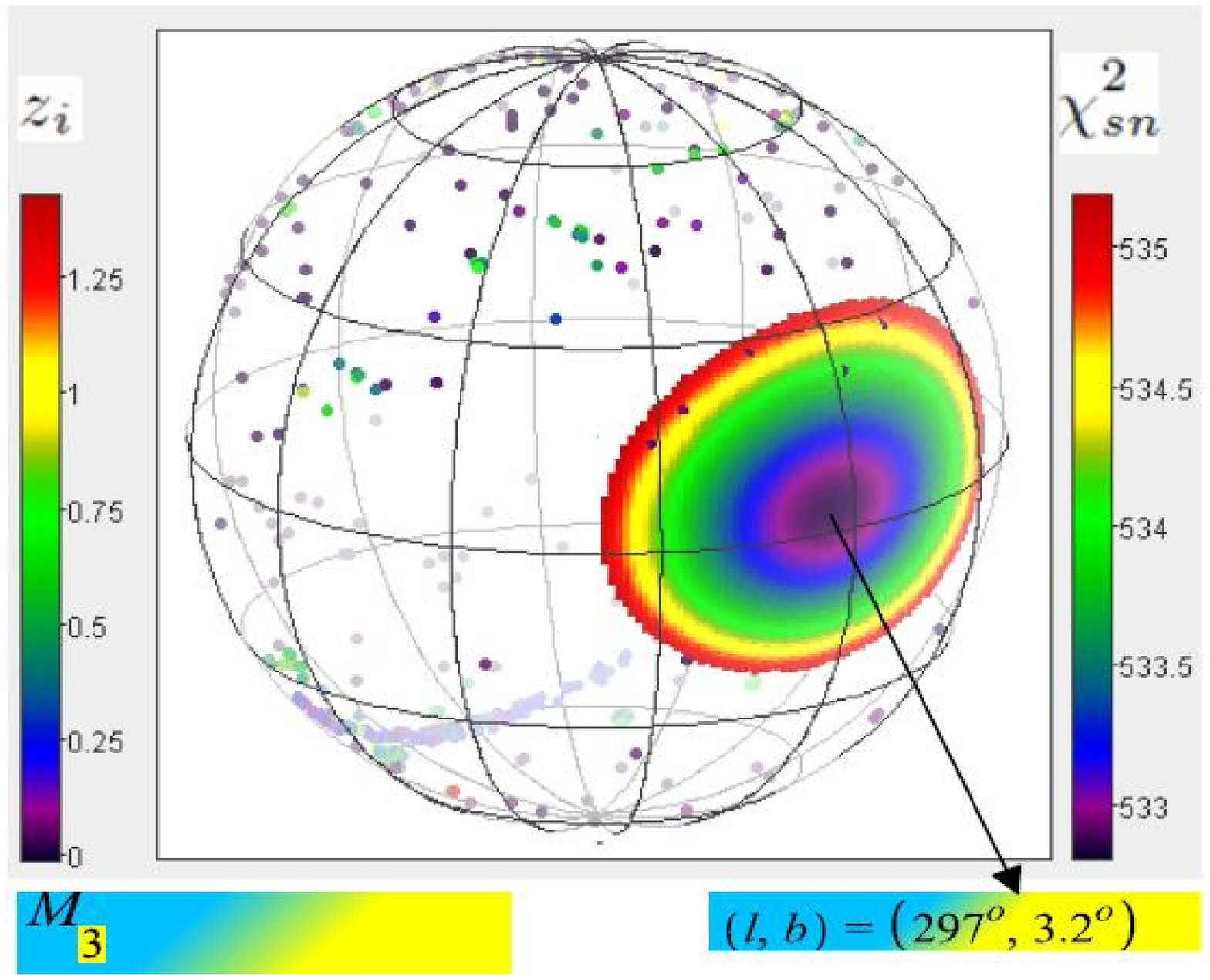}\hspace{0.1 cm}\\
Fig.8: \emph{$1-\sigma$  confidence level for parameters $(l,b)$}. \\ \emph{SNe Ia samples and the dipolar expansion direction in the galactic coordinates} \\ \emph{(used Monte Carlo simulation with} $2.5 \times10^{4}$datasets)
\end{tabular*}\\
\begin{tabular*}{2.5 cm}{cc}
\includegraphics[scale=.8]{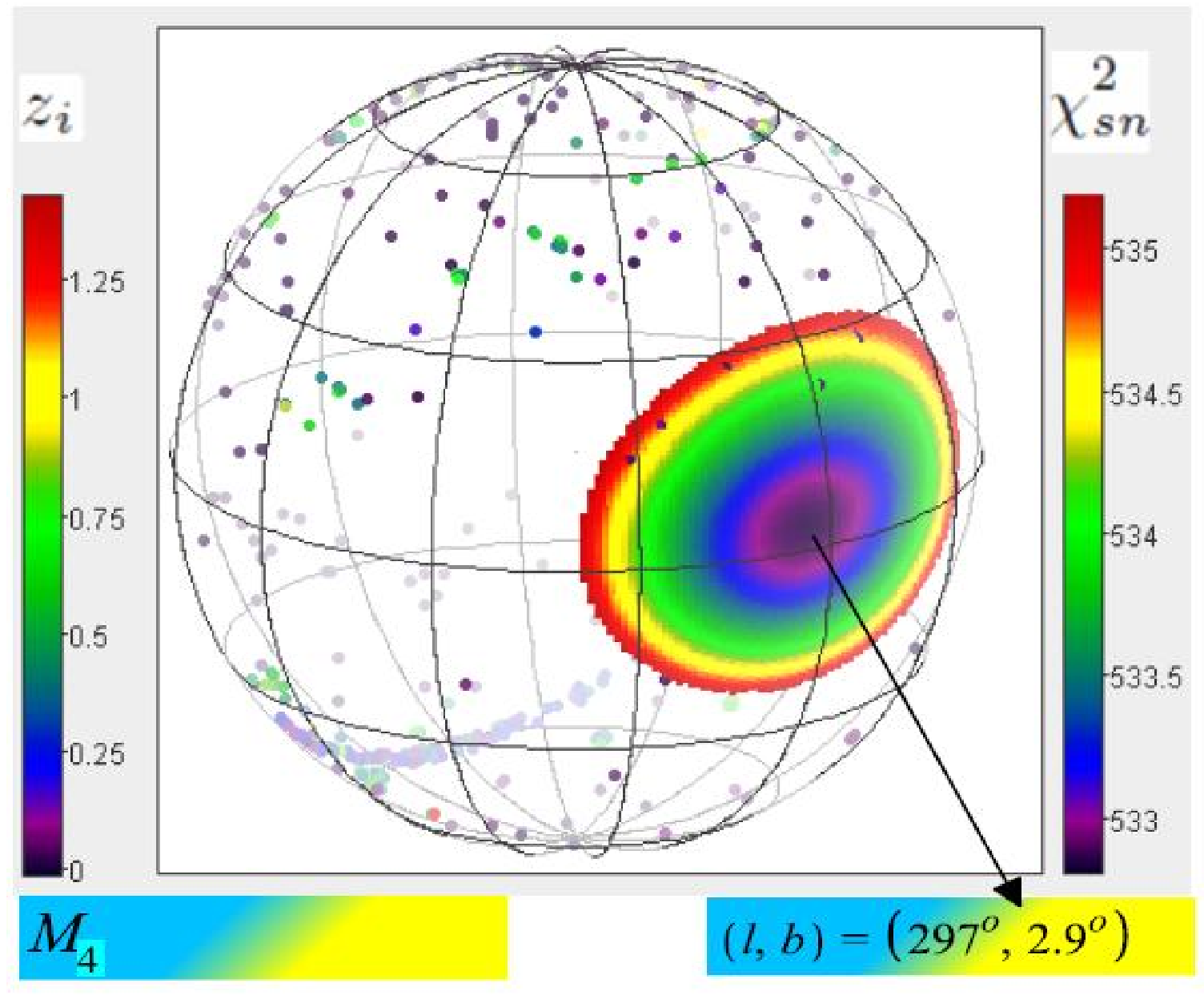}\hspace{0.1 cm}\\
Fig.9: \emph{$1-\sigma$  confidence level for parameters $(l,b)$}. \\ \emph{SNe Ia samples and the dipolar expansion direction in the galactic coordinates} \\ \emph{(used Monte Carlo simulation with} $2.5 \times10^{4}$datasets)
\end{tabular*}\\

\begin{tabular*}{2.5 cm}{cc}
\includegraphics[scale=.7]{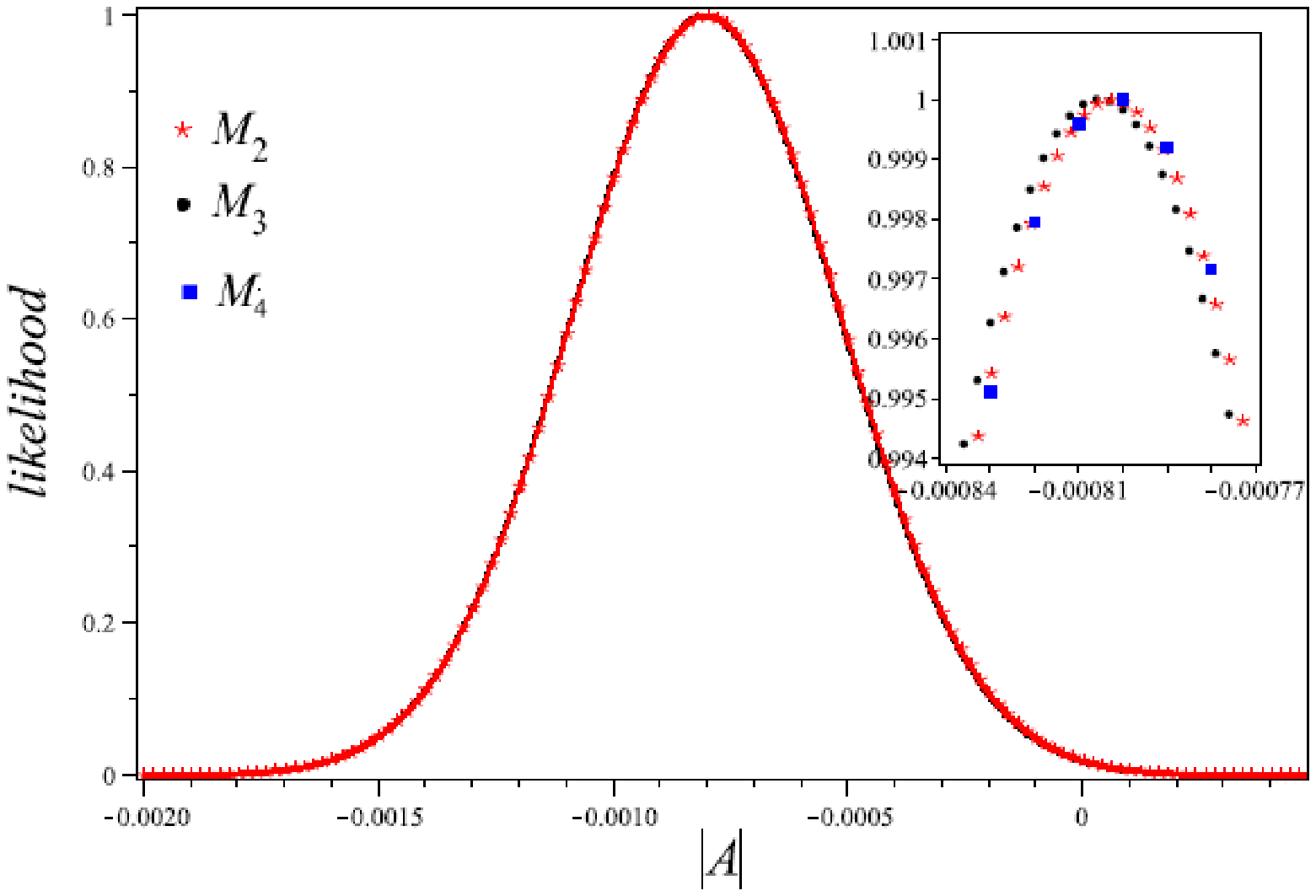}\hspace{0.1 cm}\\
Fig.10:\emph{The Anisotropic Magnitude $ A$ In The 600 Simulations Follow}\\ \emph{The Gauss Distribution}
\end{tabular*}\\

\begin{tabular*}{2.5 cm}{cc}
\includegraphics[scale=.7]{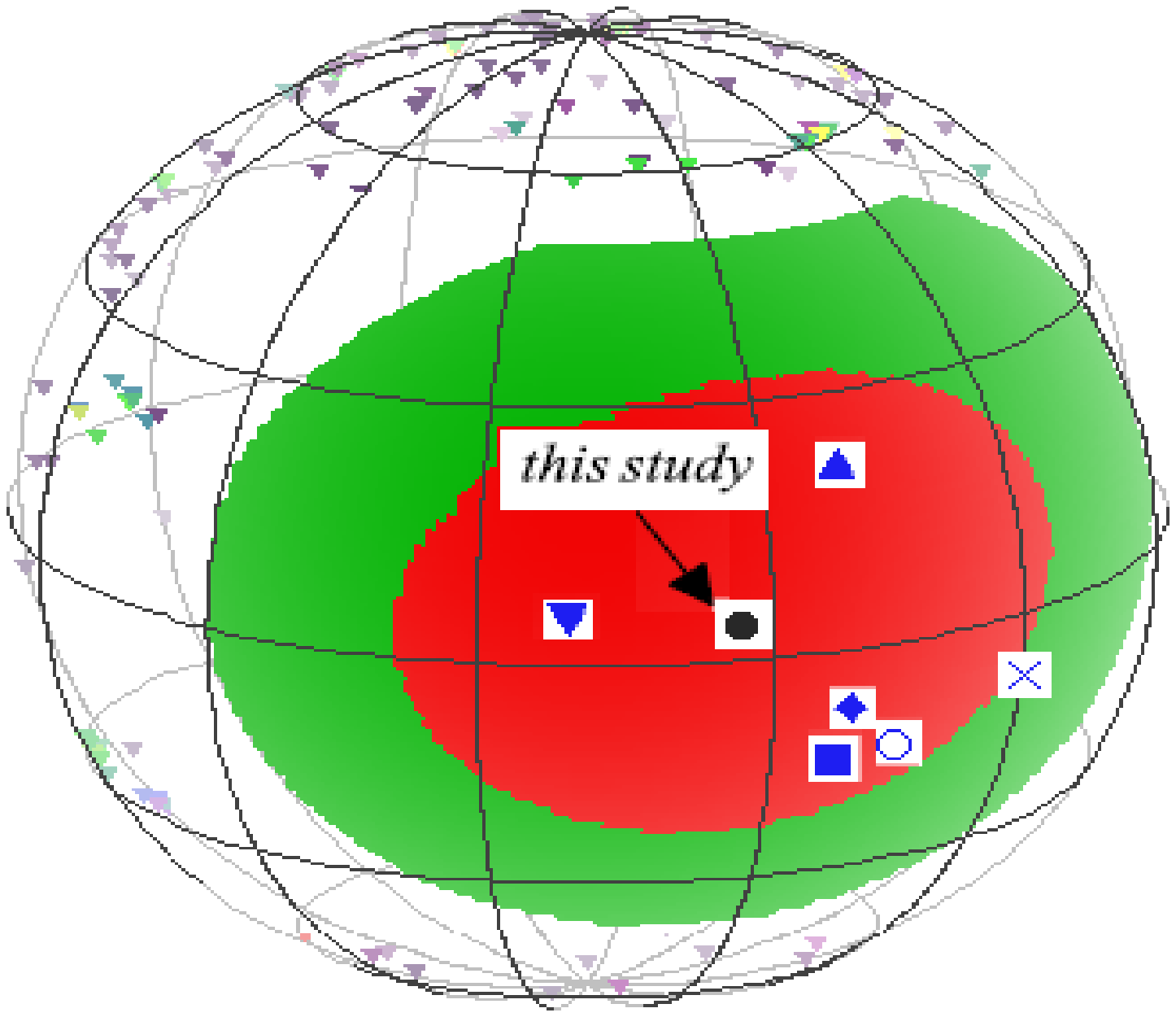}\\
Fig.11: the 1-$\sigma$and 2-$\sigma$ regions on the Dark Energy dipole direction, which include \\the results for preferred direction in other models.\\direction  of preferred axis in galactic coordinate. The point  \\ $\bullet$ denotes our result,
namely,$(l, b)=(298.7^{-32}_{+32},3^{-28}_{+28})$. The results for \\ preferred direction in other models are presented for contrast. Point $\blacktriangle$ denotes\\ the  result of \cite{Antoniou}, point $\blacksquare$ denotes the result of \cite{Xiaofeng} and \cite{a128} ,point $\blacktriangledown$ denotes the result \\of \cite{Watkins}, point $\blacklozenge$ denotes the result of \cite{Wang},\\ point $\times$ denotes the result of \cite{chang}, point $\circ$ denotes the result of \cite{chang}, \\ .
\end{tabular*}\\
\section{CONCLUSIONS AND DISCUSSIONS}
From some astronomical observations and some the
oretical models of the universe, there seemingly exists
some evidence for a cosmological preferred axis.
In this paper, we study the anisotropic expansion of the universe using type Ia
supernovae Union2 sample.The luminosity
distance is expanded with model-independent cosmographic parameters as a
function of redshift $z$ directly.The advantage of this method is that it does not rely on the particular cosmological model.By defining modified redshift$\tilde{z}$ the anisotropic luminosity distance and consequently the transformation matrix $M$ have been obtained.
 We performed statistical analysis for the $M_{1}$,$M_{2}$,$M_{3}$ and $M_{4}$ cases of Matrix $M$ ,corresponding to the  first,second,third and fourth order of Luminosity distance expansion  .We found that for $n>2$, the Model($M_{n}$) is not very sensitive to order of expansion.Thus we cut our analysis for $n>4$. We found the direction of preferred axis for $M_{2}$ as $(l,b)\simeq(297^{-34}_{+34},3.6^{-28}_{+28})$ ,$M_{3}$ as $(l,b)\simeq(297^{-33}_{+34},3.2^{-27}_{+28})$  and $M_{4}$ as $(l,b)\simeq(297^{-35}_{+34},2.9^{-28}_{+29})$ which are very close to each other. Also
 these results are compatible with other studies in $(1-\sigma)$ error region\cite{Cai0}-\cite{Watkins}.Also the magnitude of dipole in these cases are  very close to each other( $A\simeq 10^{-3}$ ) and other pervious studies.

\newpage

\def\ijmp{{\it Int. Journ. Mod. Phys.}\ }
\def\etal{{\it et al.}}
\def\prl{{\it Phys. Rev. Lett.}\ }
\def\pr{{\it Phys. Rev.}\ }
\def\modpl{{\it Mod. Phys. Lett.}\ }
\def\cqg{{\it Class. Quantum Grav.}\ }

\end{document}